\documentclass[11pt,twoside]{article}

\begin{document}
\title{\bf Quantum secret sharing with two qubit bipartite mixed state }
\author{Satyabrata Adhikari \thanks{
tapisatya@gmail.com} \\
\textsl{Department of Physics, Korea Advanced Institute of Science and Technology}\\
\textsl{Daejeon 305-701, Korea}\\
} \maketitle
\begin{abstract}
\noindent Quantum secret sharing is one of the most important and
interesting quantum information processing task. In quantum secret
sharing, information is split among several parties such that only
one of them is able to recover the qubit exactly provided all the
other parties agree to cooperate. To achieve this task, all the
parties need to share entangled state. As far as my knowledge, all
the previous quantum secret sharing protocol used either pure
tripartite or pure bipartite entangled state. In this work we use
for the first time bipartite two qubit mixed state (formed due to
noisy environment) in quantum secret sharing scheme. We further
show that one party cannot extract the information without the
collaboration of other party. We also study the property of the
shared mixed state used in the quantum secret sharing scheme.
\end{abstract}
\section{Introduction}
Quantum entanglement \cite{einstein} is one of the fascinating
feature of quantum mechanics. There is no classical analog of
quantum entanglement and that makes it more fascinating than
anything else in physics. In the field of quantum information
theory entanglement plays a major role. This is also a very useful
resource in the sense that using entanglement one can do many
things in the quantum world which are usually impossible in
ordinary classical world. Some of these tasks are quantum
computing \cite{bennett1}, quantum teleportation \cite{bennett2},
quantum cryptography \cite{gisin} and quantum secret sharing
\cite{hillery}. In quantum secret sharing, quantum information
encoded in a qubit is split among several parties such that only
one of them is able to recover the qubit exactly provided all the
other parties agree to cooperate. Therefore, quantum secret
sharing is a very interesting quantum information processing task
which was introduced in \cite{hillery}. After its introduction,
Karlsson et.al. \cite{karlsson} studied the similar quantum secret
sharing protocol using bipartite pure entangled state. Many
authors studied the concept of quantum secret sharing using
tripartite pure entangled states
\cite{bandyopadhyay,bagherinezhad,lance,gordon,zheng}. Recently Q.
Li et.al. \cite{li} proposed semi-quantum secret sharing protocols
using maximally entangled GHZ state which was shown to be secured
against eavesdropping. Quantum secret sharing can also be realized
in experiment \cite{tittel, schmid, schmid1, bogdanski}. \vskip
0.1cm

\noindent In this work we discuss the quantum secret sharing
protocol in a following way: Let us suppose that a spy (Charlie)
who is working under two commanders, Alice and Bob. Charlie's job
is to sent the secret information to both the commanders. But
Charlie suspect that one of the commanders is dishonest but he
don't know who is the culprit (Alice or Bob)? i.e. he don't know
who tries to find out the secret all by himself. So he decided to
sent the secret information in such a way that one commander
cannot collect the secret information without the help of other
commander. How Charlie achieve this task is the main result of
this work. To split information among two parties we use bipartite
mixed state. Therefore, we discuss our quantum secret sharing
protocol with two qubit mixed bipartite state. In section-2, we
review generalised concurrence and quantify the maximum amount of
entanglement present in Schmidt rank $r$ pure state in $k\times
k$-dimensional system. In section-3, we study the pure state
living in $k\times k$-dimensional Hilbert space through the noisy
environment. For two qubit system, we find that the mixed state
(because of noisy environment) shared between two distant partners
remains entangled if the concurrence of the initial entangled
state greater than certain threshold value. In section-4, we use
the two qubit mixed state (discussed in section-3) in
demonstrating the quantum secret sharing protocol. In section-5,
we end with conclusion.

\section{Generalised Concurrence - A review }
Hill and Wootters \cite{hill} introduced the first measure of
entanglement for a pair of qubits and the name given to the
entanglement measure is concurrence. For $2\times 2$- dimensional
system, the concurrence for pure state is defined as
\begin{eqnarray}
C(|\Psi_{AB}^{(2)}\rangle)=|\langle\Psi_{AB}^{(2)}|\sigma_{y}\otimes
\sigma_{y}|(\Psi_{AB}^{(2)})^{*}\rangle| \label{concurrence}
\end{eqnarray}
Since $|\Psi_{AB}^{(2)}\rangle$ is a pure $2\times 2$ bipartite
state so it can be expressed in a Schmidt-decomposition form as
\begin{eqnarray}
|\Psi_{AB}^{(2)}\rangle=\sqrt{\lambda_{1}}|00\rangle+\sqrt{\lambda_{2}}|11\rangle
\label{schmidt decomposition}
\end{eqnarray}
where $\lambda_{1},\lambda_{2}$ are schmidt coefficients and
$\lambda_{1}+\lambda_{2}=1$.\vskip 0.1cm

\noindent The concurrence (\ref{concurrence}) for the state
(\ref{schmidt decomposition}) reduces to
\begin{eqnarray}
C(|\Psi_{AB}^{(2)}\rangle)=\sqrt{2(1-Tr(\rho_{A}^{2}))}=2\sqrt{\lambda_{1}\lambda_{2}}
\label{concurrence1}
\end{eqnarray}
where
$\rho_{A}=Tr_{B}|\Psi_{AB}^{(2)}\rangle\langle\Psi_{AB}^{(2)}|$
denotes the reduced density operator.\vskip 0.1cm

\noindent Rungta et.al. \cite{rungta} then generalised the
concurrence of two-qubit pure state to higher dimensional $k\times
k$ system and the generalised concurrence (or $I-concurrence$) is
defined as
\begin{eqnarray}
C_{I}(|\Psi_{AB}^{(k)}\rangle)=\sqrt{\frac{k}{k-1}(1-Tr(\rho_{A}^{2}))}
\label{I-concurrence}
\end{eqnarray}
where
$\rho_{A}=Tr_{B}|\Psi_{AB}^{(k)}\rangle\langle\Psi_{AB}^{(k)}|$,~~
$|\Psi_{AB}^{(k)}\rangle=\sum_{i=1}^{k}\sqrt{\lambda_{i}}|i_{A}\rangle|i_{B}\rangle$,~~and
~ $\sum_{i=1}^{k}\lambda_{i}=1$.\vskip 0.1cm

\noindent $I-Concurrence$ can also be expressed in terms of
Schmidt coefficients as \cite{gour}
\begin{eqnarray}
C_{I}(|\Psi_{AB}^{(k)}\rangle)=
\sqrt{\frac{S_{2}(\lambda_{1},\lambda_{2},......,\lambda_{k})}{S_{2}(\frac{1}{k},\frac{1}{k},.....,\frac{1}{k})}}\label{I-concurrence1}
\end{eqnarray}
where $S_{2}(\lambda_{1},\lambda_{2},......,\lambda_{k})$ is the
$2nd$ elementary symmetric function of
$\lambda_{1},\lambda_{2},......,\lambda_{k}$, i.e.
$S_{2}(\lambda_{1},\lambda_{2},......,\lambda_{k})=\sum_{i<j}\lambda_{i}\lambda_{j}$.
Therefore, $I-concurrence$ can be re-written as
\begin{eqnarray}
C_{I}(|\Psi_{AB}^{(k)}\rangle)=\sqrt{\frac{2k}{k-1}\sum_{i<j,i,j=1}^{k}\lambda_{i}\lambda_{j}}
\label{I-concurrence2}
\end{eqnarray}
For $2\times 2$ dimensional system, we have
$C_{I}(|\Psi_{AB}^{(2)}\rangle)=C(|\Psi_{AB}^{(2)}\rangle)$.\vskip
0.1cm

\noindent But in reality, due to decoherence or due to preparation
error, we generally have a mixed state. Therefore, the
entanglement of the mixed state $\rho_{AB}^{(k)}=\sum
p_{i}|\Psi_{i}^{(k)}\rangle_{AB}\langle\Psi_{i}^{(k)}|$ can be
measured by convex roof extension method
\begin{eqnarray}
C_{I}(\rho_{AB}^{(k)})= min\sum_{i}
p_{i}C_{I}(|\Psi_{i}^{(k)}\rangle_{AB})\label{convexroof}
\end{eqnarray}
where the minimum is taken over all possible decomposition of
$\rho_{AB}^{(k)}$.\vskip 0.1cm

\noindent It is to be noted that the maximum amount of
entanglement in $2\times 2$-dimensional pure system is unity. They
are called maximally entangled state. Now if we proceed towards
two pure qutrit entangled systems, then we can find two SLOCC
inequivalent classes of states. The two inequivalent classes are
Schmidt rank two class (SR-2) and Schmidt rank three class (SR-3).
The pure states that belong to the Schmidt rank two class can have
amount of entanglement at most
$C_{I}(|\Psi_{2}^{(3)}\rangle)=\frac{\sqrt{3}}{2}$
($|\Psi_{2}^{(3)}\rangle$ denote the pure state of Schmidt rank 2
in $3\times 3$-dimensional system) while pure SR-3 states can
achieve the maximum amount unity. Therefore, all maximally
entangled states in two qutrit system are Schmidt rank three
(SR-3) states. Therefore, a obvious conclusion is that the amount
of entanglement in any Schmidt number 2 state in two qutrit system
is at most $\frac{\sqrt{3}}{2}$. Now it is important to ask a more
general question that if we have a $k\times k$ dimensional
entangled mixed state which has schmidt number $r$ described by a
density operator $\rho_{r}^{(k)}$ then what is the upper bound of
the amount of entanglement contained in $\rho_{r}^{(k)}$? The
answer may be given as
\begin{eqnarray}
&&(i)~~~~~C_{I}(\rho_{r}^{(k)})\leq 1,~~\textit{if~ r=k
}{}\nonumber\\&& (ii)~~~~C_{I}(\rho_{r}^{(k)})\leq
[C_{I}(|\Psi_{r}^{(k)}\rangle)]_{max},~~\textit{if~ r} <
\textit{k} \label{upperbound}
\end{eqnarray}
where $|\Psi_{r}^{(k)}\rangle$ denotes the entangled pure state of
schmidt rank r in $k\times k$-dimensional system.\vskip 0.5cm

\noindent \textbf{Theorem:} If $|\Psi_{r}^{(k)}\rangle$ denotes
the entangled pure state of schmidt rank $r$ in $k\times k$
dimensional system, then
\begin{eqnarray}
[C_{I}(|\Psi_{r}^{(k)}\rangle)]_{max}=
\sqrt{\frac{k(r-1)}{r(k-1)}} \label{maximum value}
\end{eqnarray}
\textbf{Proof:} Since $|\Psi_{r}^{(k)}\rangle$ is a entangled pure
state of schmidt rank $r$, so $|\Psi_{r}^{(k)}\rangle$ can be
expressed as
\begin{eqnarray}
|\Psi_{r}^{(k)}\rangle=\sum_{i=1}^{r}\sqrt{\lambda_{i}}|i_{A}\rangle|i_{B}\rangle,~~r=2,3,....k
\end{eqnarray}
The amount of entanglement in $|\Psi_{r}^{(k)}\rangle$ is measured
by $I-concurrence$. Therefore
\begin{eqnarray}
C_{I}(|\Psi_{r}^{(k)}\rangle)=\sqrt{\frac{2k}{k-1}\sum_{i<j,i,j=1}^{r}\lambda_{i}\lambda_{j}}
\end{eqnarray}
$C_{I}(|\Psi_{r}^{(k)}\rangle)$ can be maximized using lagrange's
multiplier method subject to the constraint
$\sum_{i=1}^{r}\lambda_{i}=1$. We find that
$C_{I}(|\Psi_{r}^{(k)}\rangle)$ attains its maximum value when
$\lambda_{1}=\lambda_{2}=........=\lambda_{r}=\frac{1}{r}$.
Therefore, the maximum value is given by
\begin{eqnarray}
[C_{I}(|\Psi_{r}^{(k)}\rangle)]_{max}=
\sqrt{\frac{k(r-1)}{r(k-1)}}
\end{eqnarray}
Hence proved.\vskip 0.1cm

\noindent \textbf{Observations:}\vskip 0.1cm

\noindent(i) If r=k, then
$[C_{I}(|\Psi_{r}^{(k)}\rangle)]_{max}=1$, as expected.\vskip
0.1cm

\noindent (ii) For higher dimensional system, i.e. as
$k\rightarrow \infty $,
$[C_{I}(|\Psi_{r}^{(k)}\rangle)]_{max}\rightarrow
\sqrt{\frac{r-1}{r}}$\vskip 0.1cm

\noindent (iii) For $k\times k$- dimensional system, we have the
following ordering of maximum value of $I-concurrence$ for
different schmidt rank states
\begin{eqnarray}
[C_{I}(|\Psi_{2}^{(k)}\rangle)]_{max}<
[C_{I}(|\Psi_{3}^{(k)}\rangle)]_{max}<[C_{I}(|\Psi_{4}^{(k)}\rangle)]_{max}<.......<[C_{I}(|\Psi_{k}^{(k)}\rangle)]_{max}=1
\label{ordering}
\end{eqnarray}
\section{Pure state through noisy environment}
In this section we study the initially prepared pure state in
$k\times k$-dimensional system passing through the noisy
environment. The state can only be used in some quantum
information processing task if it is shared between two distant
partners who wishes to exchange information between them. We
assume that Charlie is the supplier of entangled states to two
users Alice and Bob. The users of the entangled states always
demand from the supplier for the maximally entangled state. But
the supplier cannot fulfill their demand. Although supplier can
prepare maximally entangled pure state in his laboratory but the
problem is that he have to send the particles to its users through
a noisy environment. In general, the noisy environment converts
pure states to mixed states and hence the entanglement decreases
in course of distributing the particles. Due to this reason, the
users Alice and Bob have to satisfy themselves with lesser
entangled mixed state compared to pure maximally entangled
state.\vskip 0.1cm

\noindent Suppose that Charlie prepare a bipartite pure state
$|\psi\rangle^{in}$ in $k\otimes k$-dimensional system. Any
bipartite pure state can be written in the Schmidt polar form as
\begin{eqnarray}
|\psi\rangle^{in}=\sum_{i=1}^{k}\sqrt{\lambda_{i}}|i\rangle_{1}\otimes|i\rangle_{2}
\label{state}
\end{eqnarray}
where $\lambda_{i}>0,~i=1,2,........k$ are the schmidt
coefficients and satisfies the condition $\sum_{i=1}^{k}\lambda_{i}=1$.\\
After creating the entanglement between two particles, Charlie
then sent the particle 1 to Alice and particle 2 to Bob through
noisy environment. In this work, the noisy environment is
described by the unitary operator \cite{buzek}
\begin{eqnarray}
|i\rangle_{a}|0\rangle_{E}|M\rangle_{x}\rightarrow
c|i\rangle_{a}|i\rangle_{E}|X_{i}\rangle_{x}+d\sum_{j\neq
i}^{k}(|i\rangle_{a}|j\rangle_{E}+|j\rangle_{a}|i\rangle_{E})|X_{j}\rangle_{x}
\label{transformation}
\end{eqnarray}
where $|0\rangle_{E}$ denote the initial state of the environment
and $|M\rangle_{x}$ and $|X_{i}\rangle_{x} (i=1,2,....k)$ denotes
the ancilla states. The ancilla state vectors $|X_{i}\rangle_{x}
(i=1,2,....k)$ form an orthonormal basis of the ancilla Hilbert
space.\vskip 0.1cm

\noindent Unitarity of the transformation (\ref{transformation})
gives the following relation between the parameters $c$ and $d$
\begin{eqnarray}
c^{2}+2(k-1)d^{2}=1 \label{unitary}
\end{eqnarray}
When both the particles 1 and 2 is being sent through the same
noisy environment (\ref{transformation}), the state (\ref{state})
transform as
\begin{eqnarray}
|\psi\rangle^{in}\rightarrow|\psi\rangle^{out}=
c^2\sum_{i=1}^{k}\sqrt{\lambda_{i}}[|i,i\rangle_{13}\otimes|i,i\rangle_{24}|X_{i}\rangle\otimes|X_{i}\rangle]
+cd\sum_{i\neq
j}^{k}\sqrt{\lambda_{i}}|i,i\rangle_{13}\otimes{}\nonumber\\(|i,j\rangle_{24}+|j,i\rangle_{24})|X_{i}\rangle\otimes|X_{j}\rangle
+ cd\sum_{i\neq
j}^{k}\sqrt{\lambda_{i}}(|i,j\rangle_{13}+|j,i\rangle_{13})\otimes|i,i\rangle_{24}
|X_{j}\rangle\otimes|X_{i}\rangle {}\nonumber\\
+d^{2}\sum_{i=1}^{k}\sqrt{\lambda_{i}}[\sum_{i\neq
j}^{k}(|i,j\rangle_{13}+|j,i\rangle_{13})\otimes \sum_{i\neq
l}(|i,l\rangle_{24}+|l,i\rangle_{24})|X_{j}\rangle\otimes|X_{l}\rangle]
\label{output}
\end{eqnarray}
where $|\rangle_{3}$ and $|\rangle_{4}$ denote the qubit of the
environment.\vskip 0.1cm

\noindent After tracing out the ancilla qubits, four qubit state
is described by the density operator $\rho_{1324}$. When the sent
qubit 1 (2) interact with its own environment qubit 3 (4), the
state described by the density operator $\rho_{13}$ $(\rho_{24})$
can be designated as local outputs. The local output is given by
\begin{eqnarray}
&&\rho_{13}^{local}=\rho_{24}^{local}=
c^{2}\sum_{i=1}^{k}\lambda_{i}|i,i\rangle\langle i,i| +
d^{2}\sum_{i\neq
j}^{k}\lambda_{i}(|i,j\rangle+|j,i\rangle)(\langle i,j|+\langle
j,i|) \label{local}
\end{eqnarray}
Since the state described by the density operator $\rho_{14}$
$(\rho_{23})$ is formed between the sent qubit 1 (2) and
environment qubit 4 (3) located at different place so they can be
treated as non-local. The non-local output is given by
\begin{eqnarray}
\rho_{14}^{non-local}=\rho_{23}^{non-local}=P\sum_{i=1}^{k}\lambda_{i}|i,i\rangle\langle
i,i|+Q\sum_{i\neq
j}^{k}\sqrt{\lambda_{i}\lambda_{j}}|i,i\rangle\langle j,j| +
{}\nonumber\\R\sum_{i\neq j}^{k}\lambda_{i}(|i,j\rangle\langle
i,j| +|j,i\rangle\langle j,i|)+ S\sum_{l,j\neq
i}\lambda_{i}|j,l\rangle\langle j,l| \label{nonlocal}
\end{eqnarray}
where $P=(c^{2}+(k-1)d^{2})^{2}$,
$Q=d^{2}(4c^{2}+4cd(k-2)+(k-2)d^{2})$,
$R=d^{2}(c^{2}+(k-1)d^{2})$, $S=d^{4}$. \vskip 0.1cm

\noindent Alice and Bob then shared a state which is described by
the density operator $\rho_{14}$ $(\rho_{23})$. Let us now
investigate the situation for $k=2$ i.e. for two qubit
systems.\vskip 0.5cm

\noindent In the computational basis
$\{|1\rangle\otimes|1\rangle,|1\rangle\otimes|2\rangle,|2\rangle\otimes|1\rangle,|2\rangle\otimes|2\rangle\}$,
the local and non-local output is given by
\begin{eqnarray}
\rho_{13}^{local}=\rho_{24}^{local}= \left(%
\begin{array}{cccc}
  c^{2}\lambda_{1} & 0 & 0 & 0 \\
  0 & d^{2} & d^{2} & 0  \\
  0 & d^{2} & d^{2} & 0 \\
  0 & 0 & 0 & c^{2}\lambda_{2}\\
  \end{array}%
\right)
\end{eqnarray}
\begin{eqnarray}
\rho_{14}^{non-local}= \rho_{23}^{non-local}=\left(%
\begin{array}{cccc}
  P\lambda_{1}+S\lambda_{2} & 0 & 0 & Q\sqrt{\lambda_{1}\lambda_{2}} \\
  0 & R & 0 & 0  \\
  0 & 0 & R & 0 \\
  Q\sqrt{\lambda_{1}\lambda_{2}} & 0 & 0 & P\lambda_{2}+S\lambda_{1}\\
  \end{array}%
\right)
\end{eqnarray}
where $P=(c^2+d^2)^2, Q=4c^2d^2, R=c^2d^2+d^4, S=d^4$. \vskip
0.1cm

\noindent Alice and Bob shared a mixed state described by density
operator $\rho_{14}^{non-local}~~ (\rho_{23}^{non-local})$. Since
charlie sent the two particles through the noisy environment so
the state shared by the users Alice and Bob may or may not be
entangled. It depends on the noisy environment. We will find that
the shared state is entangled if there exist a critical value of
the concurrence which measures the initial entanglement present in
the two qubit pure system. This critical value of the concurrence
depends on the parameter of the noisy environment. If the
concurrence of initially prepared state less than the critical
value then the shared state is separable. We use witness operator
to find this critical value of the concurrence.\vskip 0.1cm

\noindent The optimal witness operator for two qubit system
$W_{1}^{(2)}$ is given by \cite{bertlmann}
\begin{eqnarray}
W_{1}^{(2)}=\frac{1}{2\sqrt{3}}(I-\vartheta) \label{bwitness}
\end{eqnarray}
where $\vartheta$ can be expressed in terms of the pauli matrices
$\sigma_{x},\sigma_{y}~\textrm{and}~\sigma_{z}$ as
\begin{eqnarray}
\vartheta =
\sigma_{x}\otimes\sigma_{x}-\sigma_{y}\otimes\sigma_{y}+\sigma_{z}\otimes\sigma_{z}\label{paulimat}
\end{eqnarray}
In matrix form W can be re-expressed as
\begin{eqnarray}
W_{1}^{(2)}= \left(%
\begin{array}{cccc}
  0 & 0 & 0 & \frac{-1}{\sqrt{3}} \\
  0 & \frac{1}{\sqrt{3}} & 0 & 0  \\
  0 & 0 & \frac{1}{\sqrt{3}} & 0 \\
 \frac{-1}{\sqrt{3}} & 0 & 0 & 0\\
  \end{array}%
\right)
\end{eqnarray}
Therefore,
\begin{eqnarray}
Tr(W_{1}^{(2)}\rho_{14})=Tr(W_{1}^{(2)}\rho_{23})=(\frac{-2}{\sqrt{3}})(Q\sqrt{\lambda_{1}\lambda_{2}}-R)\label{trace}
\end{eqnarray}
The non-local output $\rho_{14}^{non-local}=\rho_{23}^{non-local}$
is entangled if
\begin{eqnarray}
Q\sqrt{\lambda_{1}\lambda_{2}}-R>0\Rightarrow
2\sqrt{\lambda_{1}\lambda_{2}}=C_{I}(|\psi\rangle^{in})>C_{I}^{cr}(|\psi\rangle^{in})=\frac{1+c^2}{4c^2},~~~\frac{1}{\sqrt{3}}<c\leq1
\label{cond}
\end{eqnarray}
Therefore, the critical value of the concurrence depends on the
parameter of the noisy environment. Also we note that the function
of the parameter c is a decreasing function so the critical value
of the concurrence decreases as c increases. Thus, the lower value
of the concurrence of the initially prepared entangled state may
keep the non-local output shared state entangled if the noisy
parameter c tends towards unity.\vskip 0.1cm

\noindent It is clear that the local output state described by the
density matrix $\rho_{13}^{local}=\rho_{24}^{local}$ is separable
because
$Tr(W_{1}^{(2)}\rho_{13}^{local})=Tr(W_{1}^{(2)}\rho_{24}^{local})=\frac{1}{3\sqrt{3}}>0$.\vskip
0.1cm

\noindent We should note that the optimal witness operator
$W_{1}^{(2)}$ that detect the entangled mixed state described by
the density operator $\rho_{14}=\rho_{23}$ is not unique. There
exist another optimal witness operator \cite{sanpera} which
produce the same result (\ref{cond}) is of the form
\begin{eqnarray}
W_{2}^{(2)}=\frac{1}{2}(I-\vartheta) \label{bwitness1}
\end{eqnarray}
where $\vartheta$ is given by (\ref{paulimat}).\vskip 0.2cm

\noindent \textbf{Observation:} If Charlie initially prepare a
maximally entangled state, i.e. when
$\lambda_{1}=\lambda_{2}=\frac{1}{2}$, then for some specific
value of noisy parameter $c=\sqrt{2/3}$, the shared state between
Alice and Bob takes the form of maximally entangled mixed state.
The form of maximally entangled mixed state is given by
\begin{eqnarray}
\rho_{23}^{non-local}=\rho_{14}^{non-local}= \left(%
\begin{array}{cccc}
  \frac{13}{36} & 0 & 0 & \frac{4}{18} \\
  0 & \frac{5}{36} & 0 & 0  \\
  0 & 0 & \frac{5}{36} & 0 \\
  \frac{4}{18} & 0 & 0 & \frac{13}{36}\\
  \end{array}%
\right)=\frac{4}{9}|\Phi^{+}\rangle\langle\Phi^{+}|+\frac{5}{36}I_{4}
\label{werner}
\end{eqnarray}
where
$|\Phi^{+}\rangle=\frac{1}{\sqrt{2}}(|00\rangle+|11\rangle)$.\vskip
0.1cm

\noindent Thus if maximally entangled pure state sent through
noisy environment defined in (\ref{transformation}) then there
exist a value of the noisy parameter which transform the maximally
entangled pure state to a maximally entangled mixed state which
belongs to the family of Werner state \cite{werner}.

\section{Application of two-qubit bipartite mixed state in a quantum secret sharing problem}
In this section, we discuss a protocol for quantum secret sharing
using two-qubit bipartite mixed state. Our protocol can be
described in a few step given below: \vskip 0.5cm

\noindent \textbf{Step-I: Maximally entangled pure state prepared
by Charlie}\vskip 0.3cm

\noindent A $\textit{secret agent}$ called Charlie want to
distribute his collected confidential secret to two senior
officers called Alice and Bob in such a way that one officer
(Alice/Bob) alone cannot gather all the confidential information
by herself/himself. To accomplish his task, Charlie prepare a two
qubit maximally entangled pure state either in the form
$|\phi^{+}\rangle=\frac{1}{\sqrt{2}}(|00\rangle+|11\rangle)$ or in
the form
$|\phi^{-}\rangle=\frac{1}{\sqrt{2}}(|00\rangle-|11\rangle)$. He
would like to make his decision on $|\phi^{+}\rangle$ or
$|\phi^{-}\rangle$  by tossing a coin. If "head" appears then he
prepare $|\phi^{+}\rangle$, otherwise $|\phi^{-}\rangle$. We can
designate "head" as "0" and "tail" as "1". In this way he encode
one bit of information into the prepared state. Then he send one
qubit to Alice and another qubit to Bob through a noisy
environment defined by the unitary transformation
(\ref{transformation}). Because of the preparation strategy and
noisy environment, Alice and Bob shared a mixed state that
described either by the density operator
\begin{eqnarray}
\rho_{AB}^{+}=
\frac{P+S}{2}(|00\rangle\langle00|+|11\rangle\langle11|)+\frac{Q}{2}(|00\rangle\langle11|+|11\rangle\langle00|)+R(|01\rangle\langle01|
+|10\rangle\langle10|)\label{shared state1}
\end{eqnarray}
or by the density operator
\begin{eqnarray}
\rho_{AB}^{-}=
\frac{P+S}{2}(|00\rangle\langle00|+|11\rangle\langle11|)-\frac{Q}{2}(|00\rangle\langle11|+|11\rangle\langle00|)+R(|01\rangle\langle01|
+|10\rangle\langle10|)\label{shared state2}
\end{eqnarray}
where $P=(c^{2}+d^{2})^{2}$, $Q=4c^{2}d^{2}$,
$R=d^{2}(c^{2}+d^{2})$, $S=d^{4}$ and $c^2+2d^2=1$. \vskip 0.5cm

\noindent \textbf{Step-II: Single qubit measurement performed by
Alice}\vskip 0.3cm

\noindent Alice then perform measurement on her qubit in the
Hadamard basis
$B_{H}=\{\frac{|0\rangle+|1\rangle}{\sqrt{2}},\frac{|0\rangle-|1\rangle}{\sqrt{2}}\}$.
It is assumed that Bob also know about the measurement basis that
Alice used. The single qubit state received by Bob after
measurement depends on the outcome of the measurement. \vskip
0.1cm





\noindent (i) \textit{If the shared state is $\rho_{AB}^{+}$ and
the measurement outcome is
$\frac{|0\rangle+|1\rangle}{\sqrt{2}}$}, then
\begin{eqnarray}
\rho_{B}^{+0}&&=
\frac{1}{p}Tr_{1}[((\frac{|0\rangle+|1\rangle}{\sqrt{2}})(\frac{\langle0|+\langle1|}{\sqrt{2}})\otimes
I_{2}
)\rho_{AB}^{+}((\frac{|0\rangle+|1\rangle}{\sqrt{2}})(\frac{\langle0|+\langle1|}{\sqrt{2}})\otimes
I_{2}
)]{}\nonumber\\&&=\frac{1}{4p}[I_{2}+Q(|0\rangle\langle1|+|1\rangle\langle0|)]\label{outcome1}
\end{eqnarray}
(ii) \textit{If the shared state is $\rho_{AB}^{+}$ and the
measurement outcome is $\frac{|0\rangle-|1\rangle}{\sqrt{2}}$},
then
\begin{eqnarray}
\rho_{B}^{+1}&&=
\frac{1}{p}Tr_{1}[((\frac{|0\rangle-|1\rangle}{\sqrt{2}})(\frac{\langle0|-\langle1|}{\sqrt{2}})\otimes
I_{2}
)\rho_{AB}^{+}((\frac{|0\rangle-|1\rangle}{\sqrt{2}})(\frac{\langle0|-\langle1|}{\sqrt{2}})\otimes
I_{2}
)]{}\nonumber\\&&=\frac{1}{4p}[I_{2}-Q(|0\rangle\langle1|+|1\rangle\langle0|)]\label{outcome2}
\end{eqnarray}
(iii) \textit{If the shared state is $\rho_{AB}^{-}$ and the
measurement outcome is $\frac{|0\rangle+|1\rangle}{\sqrt{2}}$},
then
\begin{eqnarray}
\rho_{B}^{-0}&&=
\frac{1}{p}Tr_{1}[((\frac{|0\rangle-|1\rangle}{\sqrt{2}})(\frac{\langle0|-\langle1|}{\sqrt{2}})\otimes
I_{2}
)\rho_{AB}^{-}((\frac{|0\rangle-|1\rangle}{\sqrt{2}})(\frac{\langle0|-\langle1|}{\sqrt{2}})\otimes
I_{2}
)]{}\nonumber\\&&=\frac{1}{4p}[I_{2}-Q(|0\rangle\langle1|+|1\rangle\langle0|)]=\rho_{B}^{+1}\label{outcome3}
\end{eqnarray}
(iv) \textit{If the shared state is $\rho_{AB}^{-}$ and the
measurement outcome is $\frac{|0\rangle-|1\rangle}{\sqrt{2}}$},
then
\begin{eqnarray}
\rho_{B}^{-1}&&=
\frac{1}{p}Tr_{1}[((\frac{|0\rangle-|1\rangle}{\sqrt{2}})(\frac{\langle0|-\langle1|}{\sqrt{2}})\otimes
I_{2}
)\rho_{AB}^{-}((\frac{|0\rangle-|1\rangle}{\sqrt{2}})(\frac{\langle0|-\langle1|}{\sqrt{2}})\otimes
I_{2}
)]{}\nonumber\\&&=\frac{1}{4p}[I_{2}+Q(|0\rangle\langle1|+|1\rangle\langle0|)]=\rho_{B}^{+0}\label{outcome4}
\end{eqnarray}
where $I_{2}$ denotes the identity operator in $2\times
2$-dimensional Hilbert space and $p=\frac{1}{2}$.\vskip 0.3cm

\noindent(\ref{outcome3}) and (\ref{outcome4}) explains the fact
that it is neither possible for Alice nor for Bob alone to decode
the encoded information of Charlie. They only decode the
information of Charlie when they both agree to collaborate with
each other. If they agree to collaborate, then our protocol
proceeds further to step-III.\vskip 0.3cm

\noindent \textbf{Step-III: Alice declare the measurement
outcome}\vskip 0.3cm

\noindent After they agree to collaborate, Alice sent her
measurement outcome to Bob.\vskip 0.1cm

\noindent (i) If the measurement outcome is
$\frac{|0\rangle+|1\rangle}{\sqrt{2}}$ then she sent Bob a
classical bit "0" and \vskip 0.1cm

\noindent (ii) If the measurement outcome is
$\frac{|0\rangle-|1\rangle}{\sqrt{2}}$ then she sent classical bit
"1" to Bob.\vskip 0.3cm

\noindent \textbf{Step-IV: Positive operator valued measurement
(POVM) performed by Bob}\vskip 0.3cm

\noindent When Bob receives the classical bit from Alice, he came
to know about Alice's measurement outcome. Corresponding to each
measurement outcomes, one of the two possible single qubit state
may appear at Bob's site. To discriminate between the two possible
single qubit state, Bob have to perform POVM on his received
qubit. The constructed POVM at Bob's site is given by
\begin{eqnarray}
\Pi_{B}^{(0)}= \frac{1}{2}(I_{2}+\frac{1}{Q}\sigma_{x})
{}\nonumber\\\Pi_{B}^{(1)}=
\frac{1}{2}(I_{2}-\frac{1}{Q}\sigma_{x})\label{POVM}
\end{eqnarray}

\noindent If Bob receives the classical bit "0" then Alice's
measurement outcome should be
$\frac{|0\rangle+|1\rangle}{\sqrt{2}}$. Corresponding to the
Alice's measurement outcome
$\frac{|0\rangle+|1\rangle}{\sqrt{2}}$, Bob received either
$\rho_{B}^{+0}=\frac{1}{2}[I_{2}+Q(|0\rangle\langle1|+|1\rangle\langle0|)]$
or
$\rho_{B}^{-0}=\frac{1}{2}[I_{2}-Q(|0\rangle\langle1|+|1\rangle\langle0|)]$.
Bob then perform POVM to detect the correct received state. POVM
operators  $\Pi_{B}^{(0)}$ and $\Pi_{B}^{(1)}$ discriminate the
single qubit states $\rho_{B}^{+0}$ and $\rho_{B}^{-0}$ with
certainty. \vskip 0.1cm

\noindent Similarly, if Bob receives the classical bit "1" then he
can discriminate the single qubit state using POVM operators given
in (\ref{POVM}). \vskip 0.1cm

\noindent In this way our quantum secret sharing scheme work using
two qubit mixed state.

\section{Conclusion}
Before we presented our main result, we have studied generalised
concurrence or I-concurrence. We provide a compact formula to
quantify the maximum amount of entanglement present in pure state
of Schmidt rank $r$ in $k\times k$-dimensional system. We also
have studied the $k\times k$-dimensional pure state passing
through a noisy environment. We then restrict ourselves to
$2\times 2$-dimensional pure state and found that the mixed state
(because of noisy environment) shared between two distant partners
remains entangled if the concurrence of the initial entangled
state greater than certain threshold value. Thereafter, for the
first time we discussed the quantum secret sharing protocol using
two qubit mixed state which appeared due to noisy environment. Our
quantum secret sharing protocol is very simple and may be realized
in experiment. The noisy environment used in this protocol is
nothing but can be described as a quantum cloning transformation.
This type of transformation may be used by eavesdropper to steal
information. Instead of quantum cloning transformation, one may
use amplitude damping channel or any other decoherence processes.

\section{Acknowledgement}
I would like to thank Prof. H. S. Sim for discussion.

\end{document}